\documentclass{iau} 
\usepackage{graphicx,natbib,dcolumn}
\newcolumntype{d}{D{.}{.}{-1}}

\title[Braking indices] 
{Braking indices and spin evolution: \\ something is loose inside neutron stars}

\author[C.~M. Espinoza]   
{Crist\'obal M. Espinoza$^1$}

\affiliation{$^1$Departamento de F\'isica, Universidad de Santiago de Chile {\sc (usach)},\\ 
Avenida Ecuador 3493, 9170124 Estaci\'on Central, Santiago, Chile.\\ 
email: {\tt cristobal.espinoza.r@usach.cl} \\[\affilskip]}

\pubyear{2017}
\volume{337}  
\jname{Pulsar Astrophysics --­ The Next 50 Years}
\editors{P. Weltevrede, B.B.P. Perera, L. Levin Preston \& S. Sanidas, eds.}

\begin{document}

\maketitle
\begin{abstract}
Braking indices are used to describe the evolution of pulsars rotation, and can offer insights into the braking mechanism that dominates the slow down.
Here we discuss the main difficulties associated with measuring braking indices and the complexity of interpreting these measurements.
Considering recent braking index measurements on pulsars with large and regular glitches, we comment on the significant effects that the loosely coupled superfluid inside pulsars might have on their spin evolution.
\end{abstract}


The discussion presented below is based on the measurements and conclusions published by \citet{els17}.
We refer to this article for more references and details.

\section{Pulsar spin-down}
The regular spin down of pulsars was detected shortly after their discovery and remains the most prominent feature of their rotation \citep{col69,dhs69}.
As it was pointed out in early works, one evident mechanism to slow down a pulsar is magnetic braking \citep[e.g. ][]{go69}.
Assuming a dipolar configuration and considering that the magnetic axis is misaligned with the rotation axis, the system must radiate and lose energy. 
This is the one mechanism that we can be certain is participating in the slow down of all pulsars.
However, other mechanisms could also contribute.
We know that particles are being accelerated in the magnetospheres of many pulsars, and it is possible that energy loses due to a particle wind could contribute significantly to the regular spin-down \citep{hck99}.
Observable evidences for such processes are in abundance: pulsars shine, from radio to gamma-rays;  there are pulsar wind nebulae around the most energetic pulsars; and we observe spin-down switches correlated with pulse shape variations, indicative of the direct influence of magnetospheric processes on pulsars spin-down \citep[e.g.][]{klo+06,bap+17}.
However, the extent to which this could contribute to the observed evolution is unclear.
The presence and effects of other processes, such as accretion, quantum vacuum friction or the emission of gravitational waves are more uncertain and conditioned to third factors, not necessarily present in most pulsars.

\subsection{Braking index versus braking mechanism}
The braking index is used to parametrise the spin evolution of pulsars and is determined by the braking mechanisms that dominate the process.  
However, recovering such information from this number alone is complex, and the few measurements available have not offered clear information regarding the actual braking mechanism.
In general, any acting mechanism might be subject to a process of evolution at some timescale. 
For instance, the magnetic field could be evolving, either in strength or geometry, the shape of the star could be changing, or the effective moment of inertia could change due to decoupling of superfluid to the rotation of the crust. 
The observed braking index will reflect these structural or evolutionary processes if the spin-down occurs in a comparable timescale.
Therefore, a measured braking index will be unable to uniquely indicate what is the main braking mechanism behind the observed spin-down.
Even if there were no evolutionary processes involved, the case of two competing mechanisms would  also add uncertainties on the interpretation. 
Thus, braking indices must be taken with caution and to correctly interpret them we need third party information, perhaps coming from other observations.


\section{Measuring the braking index}
To define the braking index the spin evolution is assumed to follow a relation such as $\dot{\nu}\sim\nu^n$, and $n=3$ for the case of pure constant magnetic dipole braking.
The braking index can be calculated as 
\begin{equation}
n=\frac{\nu\ddot{\nu}}{\dot{\nu}^2} \quad ,
\end{equation}
provided the spin frequency $\nu$ and its first two time derivatives can be measured.
While measuring $\nu$ and $\dot{\nu}$ is easy to achieve using standard timing techniques, measuring the long-term $\ddot{\nu}$ is tremendously challenging.
One first problem comes from the fact that this is a very small quantity.
For the youngest pulsars (and for a presumed $n=3$) we should expect values $\ddot{\nu}<10^{-20}$\,Hz\,s$^{-2}$, and this limit becomes rapidly smaller for older (lower $|\dot{\nu}|$) pulsars.
In order to detect $\ddot{\nu}$, it is necessary to measure a change $\Delta\dot{\nu}$, which can be expressed in terms of the characteristic age $\tau_c=-\nu/2\dot{\nu}$ as
\begin{equation}
\frac{\Delta\dot{\nu}}{\dot{\nu}}=\frac{\Delta T}{\tau_c}\frac{n}{2} \quad,
\end{equation}
where $\Delta T$ is the duration of the observations and the timescale associated with the $\ddot{\nu}$ that is being measured.
Here, for this order of magnitude calculations we assume $\Delta T=30$\,yr and $n=3$.
Only for the Crab pulsar and others with $\tau_c\sim 1$\,kyr this quantity is larger than $10^{-2}$.
This is why these have been the first pulsars having their braking indices measured.
However, for the vast majority of pulsars this quantity is far below $10^{-3}$.
Only young pulsars such as Vela and others with $\tau_c<80$\,kyr are above this value.

These numbers must be compared with the effects of the two main rotational irregularities present in the data, which are intrinsic to most pulsars rotation, namely glitches and timing noise.
Both of them are known for introducing variations of $\dot{\nu}$ in the data, which become a problem at the time of measuring a long-term $\ddot{\nu}$.
In the case of glitches, each large event (a typical Vela-like glitch) introduces a change \citep{elsk11}
$$\frac{\Delta\dot{\nu}_\textrm{\small glitch}}{\dot{\nu}}\leq10^{-3}\quad .$$
In the case of timing noise \citep{lhk+10,hlk10}, the changes in $\dot{\nu}$ (switches) occur in a semi-periodic way, with amplitudes that can go up to 
$$\frac{\Delta\dot{\nu}_\textrm{\small tn}}{\dot{\nu}}\leq10^{-1}\quad .$$

\begin{table}
  \caption{Known braking indices between glitches ($n_\textrm{ig}$) and long-term braking indices ($n$). 
    Uncertainties (1-sigma) on the last quoted digit are shown between parentheses.}
    \begin{center}
  \label{tabla}
   {\small 
  \begin{tabular}{lddl}
  \hline
  \multicolumn{1}{l}{Name} & \multicolumn{1}{c}{$n_\textrm{\small ig}$} &  \multicolumn{1}{c}{$n$} & \multicolumn{1}{l}{References}  \\
  \multicolumn{1}{l}{} & \multicolumn{1}{c}{} &  \multicolumn{1}{c}{} & \multicolumn{1}{l}{}  \\
   \hline
   B0531$+$21 (Crab)      &  2.519(2)   & 2.342(1)  & \citet{ljg+15}    \\ 
   J0537$-$6910           & \sim20^a    & -1.22(4)  & \citet{aeka17}    \\ 
   B0540$-$69             & 2.13(1)     &  --       & \citet{fak15}     \\ 
   B0833$-$45 (Vela)      & 41.5(3)     & 1.7(2)    & \citet{els17}     \\ 
   J1119$-$6127           & 2.684(2)^b  &  --       & \cite{wje11}      \\ 
   J1208$-$6238           & 2.598(1)    &  --       & \cite{cpw+16}     \\ 
   B1509$-$58             & 2.832(3)    &  --       & \cite{lk11}       \\ 
   B1727$-$33             & --          & 1.8(3)    & \citet{els17}     \\
   J1734$-$3333           & 0.9(2)      &  --       & \cite{elk+11}     \\ 
   B1737$-$30             & --          & 1(1)      & \citet{els17}     \\
   B1757$-$24             & --          & 1.1(4)    & \citet{els17}     \\  
   B1800$-$21             & 25.9(4)     & 1.9(5)    & \citet{els17}     \\ 
   B1823$-$13             & 29.5(4)     & 2.2(6)    & \citet{els17}     \\ 
   J1833$-$1034           & 1.857(1)    &  --       & \cite{rgl12}      \\ 
   J1846$-$0258           & 2.65(1)^c   &  --       & \cite{lkg+07}     \\ 
   J2229+6114             & --          & 0(1)      & \citet{els17}     \\
\hline 
\end{tabular} } \\
\end{center}
{\sc Note.---} Not included in this compilation are braking index measurements or reported $n$ switches that are based on observations having short time spans ($<5$\,yr) compared to those used in the above measurements. \\
$^a$See \citet{aae+17} and \citet{fagk17} for a more detailed discussion of the inter-glitch behaviour of this pulsar.\\
$^b$A possible reduction of about 15\% was observed after a large glitch \citep{awe+15}. \\
$^c$Value was found to decrease to $n=2.19$ after a large glitch \citep{lnk+11,akb+15}. 
%
%
\end{table}

\subsection{Time scales on glitching pulsars}
Different timescales must be recognised at the moment of evaluating the braking index in a glitching pulsar.
As an example, let us consider the Crab pulsar \citep{ljg+15}.
This pulsar evolves with a general, long-term braking index of $n=2.342(1)$, as determined by its very large and easy-to-detect long-term $\ddot{\nu}$.
However, between glitches the same measurement is $n_\textrm{ig}=2.5$.
A similar, but more dramatic difference is seen in the Vela pulsar (and all Vela-like pulsars), where the long-term value $n=1.7$ is much smaller than the braking index measured between glitches, which in average amounts to $n_\textrm{ig}=41.5$ \citep{els17}.
These differences are caused by the persistent negative steps left by every glitch, which in the case of Vela-like pulsars can be very large.
Because there is no observed recovery from these steps, the net effect is a lower $\ddot{\nu}$ that gives an effective lower braking index.
An extreme example of this behaviour is PSR J0537$-$6910, the pulsar with the largest known glitch activity, that shows a clear negative effective trend conducive to $n=-1.2$ \citep{aeka17,fagk17}.
Hence, it is important to recognise these timescales at the moment of interpreting a braking index measurement.
In the case of the Vela pulsar, for example, while the observed long-term trend is probably consequence of the negative persistent steps at the glitches, the large inter-glitch braking indices could be  dominated by the recovery from the glitches (the slow response of the inner superfluid to the rotation change at the glitch).

\section{Recent measurements and interpretation}
We have recently carried out braking index measurements on a number glitching pulsars that exhibit regular large glitches and obtained new measurements for 6 of them, plus a re-measurument for the Vela pulsar.
The details of the methods employed can be found in \citet{els17} and the results obtained are summarised in Table~\ref{tabla}.
We also refer to Figure 6 in that work, where the movement on the $P$-$\dot{P}$ diagram of all pulsars with a measurement is indicated with an arrow.
The slope of motion on this diagram is given by $2-n$ and is perhaps the clearest possible interpretation we have today for these measurements.

The flow of pulsars on the $P$-$\dot{P}$ diagram has generally been assumed to have a slope of $-1$, corresponding to $n=3$.
However, our new measurements suggest that pulsars like Vela move horizontally, or even with a positive slope.
We only have measurements for a few of them but, considering that almost all pulsars in this area of the diagram exhibit large and regular glitches, it is natural to think that this is a general trend among these pulsars.

Most of the phenomenology associated with the rotation of Vela-like pulsars, such as low long-term braking indices and very high inter-glitch braking indices, can be understood and explained in terms of glitches and superfluid dynamics \citep[e.g. ][]{aeka17,els17}.
From this point of view, the main factor affecting the evolution of young pulsars is the loose interplay between the crust and the superfluid interior.
The rotational history of young pulsars is thus starred by a superfluid trying to catch up with the rotation of the crust.


%
%
%
%
%



\end{document}